\documentclass[12pt]{article}
\usepackage{amsmath,latexsym}
\usepackage{graphicx}
\begin{document}

 \title{\Huge Thin shell wormhole due to dyadosphere  of a charged black hole }
 \author{F.Rahaman$^*$ , M.Kalam$^{\ddag}$  and K A Rahman$^*$    }
\date{}
 \maketitle

 \begin{abstract}
To explain Gamma Ray Bursts, Ruffini argued that the event horizon
of a charged black hole is surrounded by a special region called,
the Dyadosphere where electric field exceeds the critical value
for $e^+$ $e^-$ pair production. In the present work, we construct
a thin shell wormhole by performing a thought surgery between two
dadospheres. Several physical properties of this thin shell
wormhole have been analyzed.
\end{abstract}

$ $

 %\bigskip
 %\medskip
  \footnotetext{ Pacs Nos :  04.20 Gz,04.50 + h, 04.20 Jb


 $*$Dept.of Mathematics, Jadavpur University, Kolkata-700 032,
 India:\\
 E-Mail:farook\_rahaman@yahoo.com\\
$\ddag$ Dept. of Phys. , Netaji Nagar College for Women, Regent Estate, Kolkata-700092, India.\\

    \mbox{} \hspace{.2in}}

\title{\Huge Introduction: }

To theoretical support of recent experimental evidence of gamma
ray bursts is an intriguing research area in modern astrophysics.
Some peoples believe that collapses of massive stars could be
responsible for these bursts. Recently, Ruffini and Collaborators
[1,2] have proposed an alternative explanation of gamma ray bursts
by introducing a new concept of dyadosphere of an electromagnetic
black hole. They have claimed that the event horizon of charged
black hole is encircled by a special region called dyadosphere
where the electromagnetic field strength exceeds the well known
Heisenberg - Euler critical value for the electron-positron pair
creation $\epsilon_{crit} = \frac{m_e^2c^3}{\hbar e } $ ( $m_e$
and e are mass and charge of an electron respectively ).  By
considering the dyadosphere corresponding to
Reissner-Nordstr\"{o}m spacetime, Ruffini [3] and Preparata et al
[4] have shown that the electron positron pair creation process
occurs over the entire dyadosphere outside the
Reissner-Nordstr\"{o}m horizon.

\pagebreak

They have also given a measure of total energy of
electron-positron pairs created within the dyadosphere. It is
proved that in presence of strong electromagnetic field, the
velocity of light propagation is affected by vacuum polarization
states which lead to super luminal photon propagation [ 5 ].
During the investigation of photon propagation around
Reissner-Nordstr\"{o}m  black hole, Daniels and Shore [6] have
shown that the super luminal photon propagation is possible due
the effect of one loop vacuum polarization on photon propagation.
So it is crucial to find the region where electric field exceeds
its classical limit and vacuum fluctuations take place. Recently,
Delorenci et al [7] have computed the correction for the
Reissner-Nordstr\"{o}m  metric from the first contribution of the
Euler-Heisenberg Lagrangian.

In 1989, Visser [8] had performed a thought surgery between  two
Schwarzschild black holes and glued together in such a way that no
event horizon is permitted to form. The  resultant structure leads
to a specific geometrical structure known as thin shell wormhole.
Recently, several authors have constructed thin shell wormholes by
surgically grafting of different black holes following Visser's
approach [9 - 18]. This approach is important as because the
exotic matter required for the creation of wormhole structure is
confined within the shell. Also, this novel approach gives a way
of minimizing the usage of exotic matter to construct a wormhole.

Our purpose with this paper is to present a new thin shell
wormhole whose required exotic matter could be collected from
cosmic mine i.e. from dyadosphere. According to Ruffini, the
sources  of gamma ray bursts are dyadospheres, so one can imagine,
these sources could be used by an advanced civilization to
construct and sustain a wormhole. We will discuss different
characteristics of this thin shell wormhole namely, time evolution
of
 the throat, stability, total amount of exotic matter.

The  paper is organized as follows :

In section 2, the reader is reminded about dyadosphere proposed by
Ruffini. In section 3, thin shell wormhole has been constructed
following Visser's techniques. The linearized stability analysis
is studied in section 4.  Section 5 is devoted to a brief
discussion.

\title{\Huge2. The Dyadosphere - a prelude:}

Ruffini and collaborators have proposed that there exists a region
outside the event horizon of a charged black hole, called
dyadosphere. In this region, the electromagnetic field is greater
than the Euler-Heisenberg critical value of electro-positron pair
production. According to them, this newly designed region is
responsible for the gamma ray bursts.

\pagebreak

 The simplest charged balck hole is the
Reissner-Nordstr\"{o}m black hole which is described by the line
element

\begin{equation}
               ds^2=  f(r) dt^2 - \frac{dr^2}{ f(r)} - r^2
               (d\theta^2 + \sin^2 \theta d\phi^2)
               \end{equation}
with
\begin{equation}
            f(r)= 1 - \frac{2GM}{c^2 r} + \frac{Q^2G}{c^4 r^2}
               \end{equation}

where, M and Q are mass and charge parameters.

It is known that the electric field in the Reissner-Nordstr\"{o}m
geometry is given by $ \epsilon = \frac{Q}{r^2}$  and this is
larger than $\epsilon_{crit}$ in the dyadosphere region. For the
Reissner-Nordstr\"{o}m  black hole, the dyadosphere is defined by
the radial interval $r_+ \leq r \leq r_{ds}$ where,

\begin{equation}
             r_+ =   \frac{GM}{c^2 }\left( 1 + \sqrt{ 1 - \frac{Q^2}{G
             M^2}} \right )
               \end{equation}

is the inner radius of the dyadosphere and $r_{ds}$ is the outer
radius defined by

\begin{equation}
             r_{ds} =  \sqrt{ \left(\frac{\hbar}{c m_e }\right) \left(\frac{GM}{c^2 }
             \right) \left(\frac{m_p}{ m_e }\right)
             \left(\frac{e}{q_p }\right) \left(\frac{Q}{\sqrt{G}M}\right)}
               \end{equation}

where $m_p = \sqrt{ \left(\frac{\hbar c}{G }\right)} $ and $q_p =
\sqrt{  \hbar c} $ are Planck mass and Planck charge. It is shown
that the dyadospheres exist for the charged black holes whose
masses lie within the range, $ 3.2 M_{\bigodot} < M_{dyado} <
6\times 10^5 M_{\bigodot} $. Ruffini et al [2] hava found that
total number of electron-positron pairs in the dyadosphere region
is ( in the limit, $ r_{rs} > > \frac{GM}{c^2}$ )

\begin{equation}
             N_{e^+ e^-} = \frac{Q-Q_{critical}}{e }\left[ 1 +
             \frac{(r_{ds} - r_+)}{\frac{\hbar}{m_e c }}\right]
               \end{equation}

During vacuum polarization process, the total energy of
electron-positron pairs from the static electric energy and
deposited within the dyadosphere is

\begin{equation}
             E_{dyado} = \frac {Q^2}{2 r_+ } \left( 1 -
             \frac {r_+}{r_{ds}} \right)\left[ 1 -
             \left(\frac {r_+}{r_{ds}}\right )^4 \right]
               \end{equation}

\pagebreak

 Delorenci et al [7] have computed the correction for
the Reissner-Nordstr\"{o}m metric from the first contribution of
the Euler-Heisenberg Lagrangian and obtained the following metric
as
\begin{equation}
               ds^2=  \left[1 - \frac{2M}{ r} + \frac{Q^2}{ r^2} - \frac{\sigma Q^4}{5r^6}\right]dt^2 -
               \frac{dr^2}{ \left[1 - \frac{2M}{ r} + \frac{Q^2}{ r^2} - \frac{\sigma Q^4}{5r^6}\right]} - r^2
               (d\theta^2 + \sin^2 \theta d\phi^2)
               \end{equation}
Here we use $ G = c = 1 $ and $\sigma$ is a parameter occurring
due to vacuum fluctuation effects ( i.e. the last term coming from
the one loop QED in the first order of the approximation ). One
can note that when $ \sigma \rightarrow 0$, Reissner-Nordstr\"{o}m
solution is recovered. DeLorenci [7] have also shown that the
correction term $\frac{\sigma Q^4}{5r^6}$ is of the same order of
magnitude as the classical Reissner-Nordstr\"{o}m charge term
$\frac{Q^2}{ r^2}$.

\title{\Huge3. Thought surgery and thin shell wormhole construction: }

Let us cut out two slices of region from the dyadosphere geometry
( 4 - spaces ) described by $ \Omega^\pm = ( x \mid r \leq a )  $,
where $ a\geq r_h$ ( position of event horizon of
Reissner-Nordstr\"{o}m  black hole ). Now taking two copies of the
remaining regions, $ M^\pm = ( x \mid r \geq a )  $, we paste the
two pieces together at the hypersurface $ \Sigma = \Sigma^\pm = (
x \mid r = a )  $. Thus 3 - spaces $ \Sigma  $ divides thew
spacetime into two distinct four dimensional Manifold $M^+$ (
inner spacetime ) and $M^-$ ( exterior spacetime ). Thus one gets
a geodesically complete manifold $ M = M^+ \bigcup M^- $ with a
matter shell at the surface $ r = a $ , where the throat of the
wormhole is located. This new construction implies that M is a
manifold with two asymptotically flat regions connected by the
throat. Since the boundary surface $ \Sigma  $, is a 3 - spaces,
we take the intrinsic coordinates in $\Sigma$  as $ \xi^i = (
\tau, {\theta}, \phi)$  with $\tau$ is the proper time on the
junction shell. To understand the dynamics of the wormhole, we
assume  the radius of the throat be a function of the proper time
$ a = a(\tau)$. The parametric equation for $\Sigma$ is defined by
\begin{equation}\Sigma : F(r,\tau ) = r - a(\tau)\end{equation}
The extrinsic curvature associated with the two sides of the shell
are
\begin{equation}K_{ij}^\pm =  - n_\nu^\pm\ [ \frac{\partial^2X_\nu}
{\partial \xi^i\partial \xi^j } +
 \Gamma_{\alpha\beta}^\nu \frac{\partial X^\alpha}{\partial \xi^i}
 \frac{\partial X^\beta}{\partial \xi^j }] |_\Sigma \end{equation}
where $ n_\nu^\pm\ $ are the unit normals to $\Sigma$,
\begin{equation} n_\nu^\pm =  \pm   | g^{\alpha\beta}\frac{\partial F}{\partial X^\alpha}
 \frac{\partial F}{\partial X^\beta} |^{-\frac{1}{2}} \frac{\partial F}{\partial X^\nu} \end{equation}
with $ n^\mu n_\mu = 1 $.

The intrinsic metric on $\Sigma$ is given by

\begin{equation}
               ds^2 =  - d\tau^2 + a(\tau)^2 d\Omega_2^2
               \end{equation}

 From Lanczos equation, one can obtain the surface stress
energy tensor $ S_j^i = diag ( - \sigma_s , -v_{\theta},
-v_{\phi}) $ (where $ \sigma$ is the  surface  energy density and
$ v_{\theta,\phi} $ , the surface tensions)  as
\begin{equation}
               \sigma_s =  - \frac{1}{2 \pi a} \sqrt{1-\frac{2M}{a} + \frac{Q^2}{a^2}-
               \frac { \sigma Q^4}{5 a^6} + \dot{a}^2}
               \end{equation}

\begin{equation}
              - v_{\theta} = - v_{\phi}
 = - v  =  \frac{1}{4\pi a} \frac{1 - \frac{M}{a} + \frac{2 \sigma Q^4}{5 a^6} +
 \dot{a}^2 + a \ddot{a} }{\sqrt{1-\frac{2M}{a} + \frac{Q^2}{a^2}-  \frac{ \sigma Q^4}{5 a^6} +
 \dot{a}^2}}
               \end{equation}

where over dot  means the derivative with respect to $\tau$.

Negative surface energy density in (12) implies the existence of
exotic matter at the shell. The negative signs of the tensions
mean that they are indeed pressures ($ - v_{\theta} = - v_{\phi}
 = - v  = p $). Here the radius of the shell is given by $a(\tau)$.
 For the static solution of the shell, we assume $\dot{a} = \ddot{a} = 0
 $. The surface mass of this thin shell can be defined as $ M_{shell} = 4 \pi a^2
 \sigma_s$ or

 \begin{equation}
              M_{shell} = - 2 \pi a \sqrt{1-\frac{2M}{a} + \frac{Q^2}{a^2}-
               \frac { \sigma Q^4}{5 a^6}}
               \end{equation}

Here the term M could be interpreted as the total mass of the
system i.e. total mass of the wormhole with two asymptotic regions
connected by the throat at thin shell  boundary surface $ \Sigma
$. The above equation implies,

\begin{equation}
              M = \frac{a}{2} + \frac{Q^2}{2a} -  \frac { \sigma Q^4}{10 a^5} - \frac { M_{shell}^2}{8 a}
               \end{equation}

It is interesting to note that $ M_{shell}$ is decreasing with
increases of M and this indicates that one could reduce the exotic
mass confined within the thin shell by increasing the mass of the
black hole. So the minimizing of usage of exotic matter lies on
the fact that how large Reissner-Nordstr\"{o}m  black hole we have
considered.

One can also find where the pressureless dust shell will occur.
From equation (13) , $p = 0$ implies

\begin{equation}
              h(a) \equiv 1 - \frac{M}{a} +  \frac { 2 \sigma Q^4}{5 a^6}
              = 0
               \end{equation}

For the suitable choices of parameters, the graph of the function
$h(a)$ indicates the point $a_d$ where $h(a)$ cuts $'a'$ axis (see
fig - 1 ).

\begin{figure}[htbp]
    \centering
        \includegraphics[scale=.4]{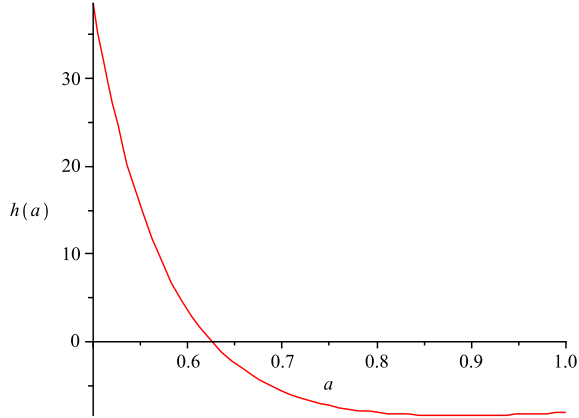}
        \caption{ Dust shell occurs at $a_d$ where $h(a)$ cuts $'a'$ axis
        ( choosing suitably the parameters
         as $ M = 10$ and $ \frac { 2 \sigma Q^4}{5} =.9 $ ).}
\end{figure}

\pagebreak

We note that the matter on the junction surface shows peculiar
behavior.  This matter violates null energy and weak energy
conditions but
  obeys    strong
energy condition.

 Here,
\begin{equation} \sigma_s < 0   \end{equation}

\begin{equation}
               \sigma_s +  p =   - \frac{\left( 1-\frac{3M}{a} +
               \frac{2Q^2}{a^2}-
               \frac { 3\sigma Q^4}{5 a^6}\right)}{4 \pi a \sqrt{1-\frac{2M}{a} + \frac{Q^2}{a^2}-
               \frac { \sigma Q^4}{5 a^6} }} < 0
               \end{equation}

\begin{equation}
               \sigma_s + 3 p =   \frac{\left( 1+\frac{M}{a} -
               \frac{2Q^2}{a^2}+
               \frac { 7\sigma Q^4}{5 a^6}\right)}{4 \pi a \sqrt{1-\frac{2M}{a} +
               \frac{Q^2}{a^2}-
               \frac { \sigma Q^4}{5 a^6} }} > 0
               \end{equation}

Now we measure the Average Null Energy Condition (ANEC) violating
matter present in the shell. This can be quantified by the
following integrals[19-20]:

\begin{equation}
             \Omega_1 =  \int \rho
\sqrt{-g}d^3x ,  \Omega_j =  \int [\rho + p_j] \sqrt{-g}d^3x
                 \end{equation}

where, $\rho = \sigma_s$, the energy condition given in (12) and
$p_j$, the principal pressures (here, radial pressure $p_r$ is
zero and transverse pressures $ p_t = p_\theta = p_\phi = - v = p
= - v_\theta = - v_\phi$ given in (13)).

\pagebreak

 Following Eiroa and Simone [11] , we introduce a new radial
coordinate $ R  =  \pm ( r -a ) $ in M ( $\pm $ for $M^{\pm}$
respectively ) so that

\begin{equation}
            \Omega_1 =  \int_0^{2\pi} \int_0^{\pi}\int_{-\infty}^\infty \rho
\sqrt{-g}dRd{\theta}d{\phi}
                 \end{equation}

\begin{equation}
            \Omega_j =  \int_0^{2\pi} \int_0^{\pi}\int_{-\infty}^\infty
            [\rho + p_j]
\sqrt{-g}dRd{\theta}d{\phi}
                 \end{equation}

Since the shell does not exert radial pressure and the energy
density is located on a thin shell surface, so that $ \rho  = \rho
+ p_r  =  \delta(R)\sigma_s$, $ \rho + p_t  = \delta(R)(\sigma_s +
p_t)$,

 Hence, one gets,
\begin{equation}          \Omega_r =   \Omega_1  = \int_0^{2\pi} \int_0^\pi
[\sigma_s \sqrt{-g} ]|_{r=a} d{\theta}d{\phi} = 4\pi a^2\sigma(a)
= -2a \sqrt{1-\frac{2M}{a} + \frac{Q^2}{a^2}-
               \frac { \sigma Q^4}{5 a^6} }\end{equation}

\begin{equation}          \Omega_t  = \int_0^{2\pi} \int_0^\pi
[(\sigma_s + p_t )\sqrt{-g} ]|_{r=a} d{\theta}d{\phi} = -
a\frac{\left( 1-\frac{3M}{a} +
               \frac{2Q^2}{a^2}-
               \frac { 3\sigma Q^4}{5 a^6}\right)}{ \sqrt{1-\frac{2M}{a} + \frac{Q^2}{a^2}-
               \frac { \sigma Q^4}{5 a^6} }}\end{equation}

one can see that if the charge of the Reissner-Nordstr\"{o}m black
hole is kept fixed, then total amount of ANEC violating matter
present in the shell is reduced by increasing the mass of the
black hole. Obviously this supports our previous analysis ( see
eq.(15) ).

\title{\Huge4. Stability Analysis: }

Rearranging equation (12), we obtain the thin shell's  equation of
motion

            \begin{equation}  \dot{a}^2 + V(a)= 0  \end{equation}

                Here  the potential is defined  as

\begin{equation}
              V(a) =  1-\frac{2M}{a} + \frac{Q^2}{a^2} - \frac{ \sigma Q^4}{5 a^6}-  4\pi^2 a^2\sigma_s^2(a)
                 \end{equation}

\pagebreak

 Linearizing around a static solution situated at $a_0$,
one can expand V(a) around $a_0$ to yield
\begin{equation}
              V =  V(a_0) + V^\prime(a_0) ( a - a_0) + \frac{1}{2} V^{\prime\prime}(a_0)
              ( a - a_0)^2 + 0[( a - a_0)^3]
                 \end{equation}
where prime denotes derivative with respect to $a$.

Since we are linearizing around a static solution at $ a = a_0 $,
we have $ V(a_0) = 0 $ and $ V^\prime(a_0)= 0 $. The stable
equilibrium configurations correspond to the condition $
V^{\prime\prime}(a_0)> 0 $. Now we define a parameter $\beta$,
which is interpreted as the speed of sound, by the relation [7]
\begin{equation}
              \beta^2(\sigma_s) = \frac{ \partial p}{\partial
              \sigma_s}|_{\sigma_s}
                 \end{equation}
Here,
\begin{equation} V^{\prime\prime}(a) = -\frac{4M}{a^3} + \frac{6Q^2}{a^4} - \frac{42 \sigma Q^4}{5a^8} - 8\pi^2 \sigma_s^2
  - 32\pi^2 a \sigma_s \sigma_s^\prime   - 8\pi^2 a^2 (\sigma_s^{\prime})^2
 - 8\pi^2 a^2\sigma_s \sigma_s^{\prime\prime}
 \end{equation}

 From equations (12) and (13), one can write energy conservation
 equation as
\begin{equation}
               \dot{\sigma_s} +  2\frac{\dot{a}}{a}( p + \sigma_s ) = 0
               \end{equation}
               or
 \begin{equation}
               \frac {d}{d \tau} (4 \pi \sigma_s a^2) + p \frac{d}{d \tau}(4\pi a^2)= 0
               \end{equation}

From equation (30) ( by using (28) ),  we obtain,

\begin{equation}
               \sigma_s^{\prime\prime} +  \frac{2}{a} \sigma_s ^{\prime} ( 1 + \beta^2)
               - \frac{2}{a^2} ( p + \sigma_s) = 0
               \end{equation}

 The wormhole solution is stable if $
V^{\prime\prime}(a_0)> 0 $ i.e. if
\begin{equation} \beta_0^2 < \frac{\left( \frac{3}{2}-\frac{6M}{a_0}
+
               \frac{5Q^2}{a_0^2} -
               \frac { 18\sigma Q^4}{5 a_0^6}\right)}{1-\frac{2M}{a} +
               \frac{Q^2}{a^2}-
               \frac { \sigma Q^4}{5 a^6} } - \frac{\left( 1-\frac{3M}{a_0}
+
               \frac{2Q^2}{a_0^2} -
               \frac { 4\sigma Q^4}{5 a_0^6}\right)}{2 \left( 1-\frac{2M}{a} +
               \frac{Q^2}{a^2}-
               \frac { \sigma Q^4}{5 a^6}\right) }-\frac{3}{2} \end{equation}
If one treats $a_0$, M ,  Q  and $\sigma $ are specified
quantities, then the stability of the configuration requires the
above restriction on the parameter $\beta_0$. This means there
exists some part of the parameter space where the throat location
is stable. For a lot of useful information, we show the stability
regions graphically ( see figure 2).

\begin{figure}[htbp]
    \centering
        \includegraphics[scale=.4]{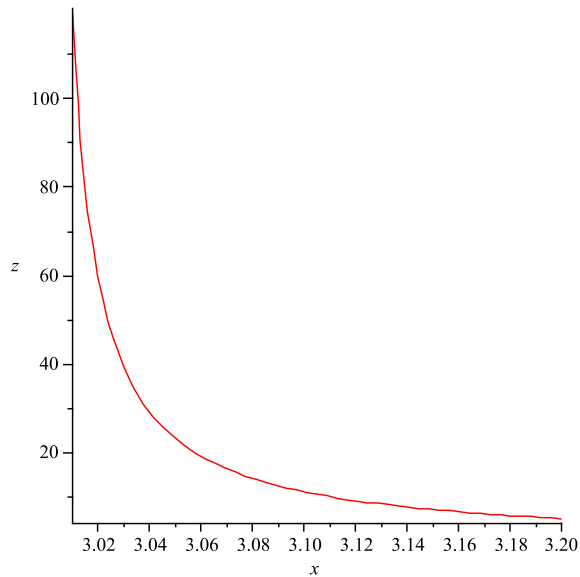}
        \caption{ Here we plot $ z = \beta^2_{|{(a=a_0)}} $ $ Vs.$ $ x= \frac{a_0}{M}$ ( choosing suitably
         the parameters as $  (\frac{Q}{a_0})^2 = .12 $ and
        $\frac { \sigma Q^4}{5 a_0^6} = .1$ ).
         The stability region is situated below the curve.}
    \label{fig:stability}
\end{figure}

\pagebreak

\title{\Huge5.  Discussions: }

The minimizing of usage exotic matter needed to construct a
wormhole remains an encouraging research area to the scientists
working in wormhole physics. Several models and ideas are proposed
time to time. Recent observations of gamma ray bursts confirmed
that there should exist some sources that produce these bursts.
Ruffini and his collaborators argued that these sources are
nothing but the dyadospheres. In this work, we have considered a
dyadosphere of  Reissner-Nordstr\"{o}m  black hole to develop thin
shell wormhole. We have constructed thin shell wormhole by
surgically grafting two dyadosphere spacetimes. This study is
interesting and more physical since we have used the spacetimes of
the astrophysical sources of cosmic gamma ray bursts. We have
analyzed the dynamical stability of thin shell wormhole
considering linearized radial perturbation around the stable
solution. We have shown that there exist some part of the
parametric space where the thin shell wormhole is stable. We have
also discussed the stability graphically. The most important part
of this study is the minimizing of usage of exotic matter confined
within the shell. We have seen that Reissner-Nordstr\"{o}m  black
hole mass plays a crucial role to minimize the usage of exotic
matter. The Reissner-Nordstr\"{o}m  black hole with larger mass
reduces the amount of exotic matter needed for their construction.
Finally, we note that the parameter $\sigma$ coming from the one
loop QED in the first order of the approximation is also
responsible for reducing the exotic matter.

\pagebreak

 {  \Huge Acknowledgments }

          F.R. is thankful to  DST , Government of India for providing
          financial support.
          \\

%\begin{figure}[p]
%\includegraphics*[450,350]{fig1.bmp}
%\caption{Variation of deflection of the circular plate}
%\end{figure}

\end{document}